\documentclass[aps,pra,preprint,superscriptaddress,showpacs]{revtex4}
\usepackage{bm}
\usepackage{amsmath}
\usepackage{amssymb}
\usepackage{mathrsfs}
\usepackage[dvips]{graphicx}
\begin{document}
\title{Effect of a strong laser field on $\bm{e}^{\bm{+}}\text{-}\bm{e}^{\bm{-}}$ photoproduction by relativistic nuclei}

\author{A. Di Piazza}
\email{dipiazza@mpi-hd.mpg.de}
\affiliation{Max-Planck-Institut f\"ur Kernphysik, Saupfercheckweg 1, 69117 Heidelberg, Germany}
\author{E. L{\"o}tstedt}
\email{lotstedt@chem.s.u-tokyo.ac.jp}
\affiliation{Max-Planck-Institut f\"ur Kernphysik, Saupfercheckweg 1, 69117 Heidelberg, Germany}
\affiliation{Department of Chemistry, School of Science, The University of Tokyo, 7-3-1 Hongo, Bunkyo-ku, Tokyo 113-0033, Japan}

\author{A. I. Milstein}
\email{milstein@inp.nsk.su}
\affiliation{Max-Planck-Institut f\"ur Kernphysik, Saupfercheckweg 1, 69117 Heidelberg, Germany}
\affiliation{Budker Institute of Nuclear Physics, 630090 Novosibirsk, Russia}
\author{C. H. Keitel}
\email{keitel@mpi-hd.mpg.de}
\affiliation{Max-Planck-Institut f\"ur Kernphysik, Saupfercheckweg 1, 69117 Heidelberg, Germany}

\begin{abstract}
We study the influence of a strong laser field on the Bethe-Heitler photoproduction process by a relativistic nucleus. The laser field propagates in the same direction as the incoming high-energy photon and it is taken into account exactly in the calculations. Two cases are considered in detail. In the first case, the energy of the incoming photon in the nucleus rest frame is much larger than the electron's rest energy. The presence of the laser field may significantly suppress the photoproduction rate at soon available values of laser parameters. In the second case, the energy of the incoming photon in the rest frame of the nucleus is less than and close to the electron-positron pair production threshold. The presence of the laser field allows for the pair production process and the obtained electron-positron rate is much larger than in the presence of only the laser and the nuclear field. In both cases we have observed a strong dependence of the rate on the mutual polarization of the laser field and of the high-energy photon and the most favorable configuration is with laser field and high-energy photon linearly polarized in the same direction. The effects discussed are in principle measurable with presently available proton accelerators and laser technology.
\end{abstract}

\pacs{12.20.Ds, 42.62.-b}

\maketitle

\section{Introduction}
A high-energy photon propagating in the Coulomb field of a nucleus may 
produce an electron-positron pair, provided the energy $\omega^*_1$ of the photon 
in the rest frame of the nucleus exceeds the threshold $2m$, with $m$ being the electron mass (units with $\hbar=c=1$ are used). This process is commonly referred to 
as the Bethe-Heitler process \cite{BeHe1934}. The well-known formula for the total cross section, valid for high energies $\omega^*_1\gg m$ reads
\begin{equation}\label{eq:BHformula}
\sigma_{BH}=\frac{28}{9}Z^2\frac{\alpha^3}{m^2}
\left[\ln\left(\frac{2\omega^*_1}{m}\right)-\frac{109}{42}\right],
\end{equation}
where $Z$ is the nuclear charge number and $\alpha$ is the fine-structure constant. Equation \eqref{eq:BHformula}
was obtained in the Born approximation with respect to the Coulomb field. In the case of a strong nuclear field 
where the parameter $Z\alpha$ is of the order of unity, high-order terms in $Z\alpha$ should be taken into
account. In some cases the inclusion of high-order terms in $Z\alpha$ essentially modifies the value of the cross section
as compared to the Born result. This was convincingly demonstrated in the case of photon scattering by a 
Coulomb field, the so-called Delbr\"uck scattering 
\cite{Milstein1994183,AkhKezetal1998}, and for Coulomb field-induced 
photon splitting \cite{AkhKezetal2002,Leeetal2003}. The cross section of the
Bethe-Heitler process exact in $Z\alpha$ has also been extensively studied
\cite{BethMax1954,OvMoOl1973,LeeMilStr2004,LeeMilStr2005}, as well as the process of $e^+\text{-}e^-$ 
production in heavy ion-collisions \cite{BaHeTra2007,Baltz20081}.

Strong electromagnetic fields are also produced in the laboratory by powerful lasers. In this case the parameter that characterizes the strength of the field is $\xi=eE/m\omega_0$ where $e$ is the absolute value of the electron charge, $E$ is the peak electric field of the laser, and $\omega_0$ is the laser frequency. Already at laser intensities of the order of $10^{18}$ W/cm$^2$ (at $\omega_0\approx 1$ eV) the parameter $\xi$ becomes of the order of unity, and the interaction with the laser field cannot be treated perturbatively. A large value of $\xi$ is in general not sufficient to have noticeable nonlinear quantum electrodynamic (QED) effects in a laser field. It is also necessary that $E$ is comparable with the so-called critical field of QED
$E_{cr}=m^2/e$, corresponding to a laser intensity of $I_{cr}=2.3\times 10^{29}\;\text{W/cm$^2$}$. Nonlinear QED effects in strong laser fields have been theoretically studied in numerous processes such as $e^+\text{-}e^-$ pair creation in photon-laser collisions \cite{Re1962,NiRi1964}, lepton production in laser-nucleus collisions \cite{Mueller_2003,MiMuHaJeKe2006,KuRo2007,Deneke_2008,Mueller_2009} also in the presence of an additional incoming photon \cite{Loetstedt_2008} and in counterpropagating laser pulses \cite{BuNaMuPo2006,bell:200403,Rufetal2009}, in particular ultrashort \cite{Hebenstreit_2009}. Other
interesting processes are laser photon merging \cite{DiPiHaKe2007}, high-energy photon splitting 
\cite{DiPiMiKe2007}, and the related laser-Coulomb Delbr\"uck scattering \cite{DiPiMi2008}. For further references to nonlinear QED effects in laser fields, see the reviews \cite{MoTaBu2006,Marklund_2006,SaHuHaKe2006}. To date, the only direct experimental verification of  $e^+\text{-}e^-$ pair creation by laser photons is reported in \cite{Buetal1997,BaBoetal2004}. Electron-positron pair production by high-energy photons, radiated by laser-accelerated fast electrons, when colliding with a nucleus has been observed several times \cite{CowPeretal1999,gahnetal2000}, most recently in \cite{CheScoetal2009}.

In the present paper, we investigate the effects of a strong laser field on the Bethe-Heitler process with the incoming photon and the laser field propagating in the same direction. We consider two cases. In the first one the energy $\omega^*_1$ in the rest frame of the Coulomb center is much larger than $m$ and the frequency of the laser field $\omega^*_0$ in the same frame is much smaller than $m$. In this case, the photon with energy $\omega_1$ (in the laboratory frame) alone can create an electron-positron pair. Also, electron-positron pairs can be produced in the superposition of laser and Coulomb fields even if the laser frequency $\omega_0$ (in the laboratory frame) is much smaller than $m$ (see for example \cite{Yakovlev_1966}). However, the kinematics of this process is different of that of electron-positron production by a high-energy photon. Therefore, it is possible to distinguish experimentally the two processes. In the second case we consider photon energies below but close to the threshold: $0<2m-\omega^*_1\ll m$, again $\omega^*_0\ll m$ and $E^*\ll E_{cr}$ ($E^*$ indicates the amplitude of the laser field in the rest frame of the nucleus) \cite{Di_Piazza_2009}. This case is interesting because the high-energy photon alone cannot produce a pair and the probability of pair creation in combined Coulomb and laser fields is very small. However, we show that the probability of photoproduction in the presence of a laser field is much larger than the probability in the case of laser and Coulomb fields (the model process of electron positron pair creation in the superposition of a strong, low-frequency electric field and a weak, high-frequency electric field has been recently investigated in \cite{Schuetzhold_2008}). In both high-energy and near-to-the-threshold cases we study the effect of the mutual polarization of the photon and of the laser field and it turns out that this effect is also important. It will be seen that the Lorentz-invariant parameter that characterizes the magnitude of the effect of the laser field is $\chi=\xi\omega_0/\omega_1=(E/E_{cr})(m/\omega_1)$. Nowadays, for the strongest available laser we have $E/E_{cr}=3\times 10^{-4}$ \cite{Emax}. Therefore, for $\omega_1\gg m$ the parameter $\chi$ will be too small and the effect negligible. However, it is possible to change the situation by using a relativistic nucleus with Lorentz factor $\gamma$ moving in the opposite direction of the photon momentum. Then, the frequency of the photon and of the laser field in the rest frame of the nucleus will be $\omega^*_{0,1}=2\gamma\omega_{0,1}$. Therefore, for large $\gamma$ it can be that $\omega^*_1\gg m$ even if $\omega_1\ll m$ and it is possible to obtain values of $\chi$ of the order of unity even for subcritical electric fields in the laboratory frame. As a result, in the rest frame of the nucleus the electric field amplitude of the laser field will be larger than $E_{cr}$. Thus, such scheme of experiment allows one to investigate nonlinear QED effects in overcritical electromagnetic fields.

The paper is organized as follows. In Section II we derive the general expression of the total pair production rate for a plane wave of arbitrary
spectral content and polarization. The case of a bichromatic field, consisting of a strong, low-frequency wave and a weak, high-frequency field is considered in Section III in the case $\omega^*_1\gg m$. The other case with $0<2m-\omega^*_1\ll m$ is investigated in Section IV. Finally, in Section V the main conclusions of the paper are presented.

\section{General expression of the total electron-positron photoproduction rate}

In this paper we calculate the total electron-positron pair production rate $\dot{W}$ in the Born approximation with respect to the Coulomb field by using the dispersion relation, which allows one to express $\dot{W}$ via the imaginary part of the polarization operator $\Pi^{\mu\nu}$ of a virtual photon in a laser field (the polarization operator approach was suggested and employed for the case of a monochromatic plane wave in \cite{MiMuHaJeKe2006}):
\begin{equation}
\dot{W}=\frac{(4\pi Ze)^2}{4\pi}\int\frac{d^3q}{(2\pi)^3}\,\frac{\text{Im}\Pi^{00}}{|\mathbf{q}|^4}.
\end{equation}
We also use the expression of the polarization operator obtained in \cite{BaMiSt1975} by means of the operator technique (the polarization operator in another form was obtained independently in \cite{BeMi1975}). In the general case the incoming electromagnetic field is described by a plane wave with vector potential $\mathbf{A}(\phi)=\mathbf{a}_1\psi_1(\phi)+\mathbf{a}_2\psi_2(\phi)$, where $\psi_i(\phi)$ with $i=1,\,2$ are two arbitrary functions of $\phi=t-z$ (the plane wave is assumed to propagate in the positive $z$ direction), and $\mathbf{a}_i$ are the two polarization vectors such that $\mathbf{a}_i\cdot\mathbf{a}_j=0$ if $i\neq j$ and $\mathbf{a}_i\cdot\hat{\mathbf{z}}=0$. By employing the expression of $\Pi^{00}$ from \cite{BaMiSt1975}, we arrive at
\begin{equation}
\label{W_dot_i}
\begin{split}
\dot{W}&=-\frac{(Z\alpha)^2m}{\pi^2}\text{Im}\int_0^{\infty}\frac{dQ}{Q^2}\int_0^1dv\int_0^{\infty}\frac{d\tau}{\tau}\int_0^T\frac{d\phi}{T}\int_0^1dx\frac{1-x^2}{x^2}\\
&\times\exp\left\{-i\tau[1+Q^2(1-v^2)+\beta]\right\}\\
&\times\left[\frac{3-v^2}{1-v^2}\mathbf{\Delta}(1)\cdot\mathbf{\Gamma}-\mathbf{\Delta}^2(1)-1-Q^2\frac{1-3v^2+x^2(1+v^2)}{1-x^2}\right],
\end{split}
\end{equation}
where the integration $\int_0^T d\phi/T$ corresponds to the average of the expression over $\phi$, $Q=|\mathbf{q}|/2m$ and $x=\mathbf{q}\cdot\hat{\mathbf{z}}/|\mathbf{q}|$. Also, the following quantities have been introduced:
\begin{align}
\label{Not_1}
\mathbf{\Delta}(y)=\frac{e}{m}[\mathbf{A}(\phi-\nu y)-\mathbf{A}(\phi)], && \mathbf{\Gamma}=\int_0^1dy\mathbf{\Delta}(y),&& \beta=\mathbf{\Gamma}^2-\int_0^1dy\mathbf{\Delta}^2(y)
\end{align}
with $\nu=Qx(1-v^2)\tau/2m$. Eq. (\ref{W_dot_i}) is the starting point of our considerations.

\section{High-energy photoproduction}

It is convenient to consider the process in the rest frame of the nucleus. For the sake of simplicity of notation we omit the sign $^*$ for all physical quantities in this frame. Let the plane electromagnetic field consist of a strong, monochromatic, low-frequency ($\omega_0\ll m$), field with adimensional vector potential components $\xi_{1,2}=eE_{1,2}/m\omega_0$ and a weak, monochromatic, high-frequency ($\omega_1\gg m$) field with adimensional vector potential components $w_{1,2}=e\mathcal{E}_{1,2}/m\omega_1$. Here $E_{1,2}$ and $\mathcal{E}_{1,2}$ are the electric field components of the strong and the weak field, respectively. Note that the strong and the weak fields propagate along the same direction. We assume that $w_{1,2}\ll 1$ which is reasonable for any available sources of high-energy photons. Let us represent the total rate $\dot{W}$ as the sum $\dot{W}_0+\dot{W}_1$, where $\dot{W}_0$ is independent of the parameters of the weak, high-frequency field and the expansion of $\dot{W}_1$ with respect to $w_{1,2}$ starts with quadratic terms. The leading contribution to the rate $\dot{W}_1$ comes from the region of integration $Qx\ll 1$, which corresponds to small longitudinal momentum transfer in comparison with the electron mass and in this limit the term proportional to $Q^2$ in the pre-exponent of Eq. (\ref{W_dot_i}) is negligible. The integrals over $Q$, $x$ and $v$ can be taken by using the condition $a=2m/\omega_1\ll 1$ and the result with logarithmic accuracy is
\begin{equation}
\begin{split}
\dot{W}_1&=\frac{2}{3}\frac{(Z\alpha)^2m}{\pi^2a}\int_0^{\infty}\frac{d\rho}{\rho^2}\int_0^T\frac{d\phi}{T}\left\langle\frac{1}{\mathcal{Z}}\left\{\left[\ln\left(\frac{2}{a\rho\sqrt{\mathcal{Z}}}\right)-C-\frac{1}{2}\right]\left[4\mathbf{\Delta}(1)\cdot\mathbf{\Gamma}-\mathbf{\Delta}^2(1)-1\right]\right.\right.\\
&\left.\left.-\frac{23}{6}\mathbf{\Delta}(1)\cdot\mathbf{\Gamma}+\frac{5}{6}\mathbf{\Delta}^2(1)+\frac{5}{6}\right\}+\ln\left(\frac{2}{a\rho}\right)-C-\frac{4}{3}\right\rangle,
\end{split}
\end{equation}
where $C=0.577...$ is the Euler constant, $\mathcal{Z}=1+\beta$ and where the variable $\rho=\omega_1\nu=Qx(1-v^2)\tau/a$ has been introduced. In this equation it is assumed that the terms independent of $w_{1,2}$, which correspond to pair production only in combined laser and Coulomb fields, are subtracted. It is worth observing that the above relatively compact expression holds for arbitrary polarization and spectral content of the low-frequency strong laser field.

By expanding the total rate $\dot{W}_1$ for small $w_{1,2}$ up to quadratic terms and by taking into account exactly the strong laser field, we obtain for monochromatic strong and weak fields
\begin{equation}
\label{W_dot}
\begin{split}
\dot{W}_1&=\frac{(Z\alpha)^2\omega_1}{3\pi^2}\int_0^{\infty}\frac{d\rho}{\rho^2}\int_0^{2\pi}\frac{d\psi}{2\pi}\frac{1}{\mathcal{Z}_0}\left\{\left[\ln\left(\frac{\omega_1}{m\rho\sqrt{\mathcal{Z}_0}}\right)-C\right]\right.\\
&\times\left[4\frac{w_1^2+w_2^2}{2}\sin^2\rho-2F_3+\left(G-\frac{F_2}{2}\right)(1-4F_1)\right]\\
&\left.-\frac{19}{3}\frac{w_1^2+w_2^2}{2}\sin^2\rho+\frac{13}{6}F_3+\frac{5}{6}G\left(\frac{26}{5}F_1-1\right)-\frac{7}{24}F_2\left(\frac{40}{7}F_1-1\right)\right\},
\end{split}
\end{equation}
where
\begin{eqnarray}
&& F_1=\sin^2 (b\rho)\left(\frac{\xi_1^2+\xi_2^2}{2}-\frac{\xi_1^2-\xi_2^2}{2}\cos\psi\right),\nonumber\\
&& F_2=\frac{1}{\mathcal{Z}_0^2}\left(F_+^2M_+^2+F_-^2M_-^2+2F_-F_+M_-M_+\cos\psi\right),\nonumber\\
&& F_3=\frac{1}{\mathcal{Z}_0}\sin(b\rho)\sin\rho\left[F_+M_+^2-F_-M_-^2-(F_+-F_-)M_-M_+\cos\psi\right],\nonumber\\
&& G=\frac{1}{\mathcal{Z}_0}\left(1-\frac{\sin^2\rho}{\rho^2}\right)\frac{w_1^2+w_2^2}{2},\nonumber\\
\label{Z_0_1}
&& \mathcal{Z}_0=1+\frac{\xi_1^2+\xi_2^2}{2}\left[1-\frac{\sin^2(b\rho)}{(b\rho)^2}\right]-\left(\frac{\xi_1^2-\xi_2^2}{2}\right)\frac{\sin(b\rho)}{b\rho}\left[\frac{\sin(b\rho)}{b\rho}-\cos(b\rho)\right]\cos\psi.
\end{eqnarray}
In these expressions we have introduced the following notation:
\begin{align}
M_{\pm}&=\xi_1w_1\pm\xi_2w_2, \qquad F_{\pm}=\frac{\sin((1\mp b)\rho)}{(1\mp b)\rho}-\frac{\sin(b\rho)}{b\rho}\frac{\sin\rho}{\rho}, \qquad b=\frac{\omega_0}{\omega_1}.
\end{align}
We note that the functions $F_2$ and $F_3$ depend on the mutual polarization of the weak and the strong field. Also, the coefficients of the logarithms in the above result in Eq. (\ref{W_dot}) can in principle be obtained by applying the Weizsacker-Williams' method of virtual photons \cite{Landau_b_4}. However, the constant terms present in Eq. (\ref{W_dot}) cannot be obtained by that method and they give, unless for initial photon energies largely exceeding $m$, quantitatively important results (see also the numerical example in Par. III.A).

For $\omega_0\ll m$ and $\omega_1\gg m$ the effect of the laser field is noticeable only if $\xi=\sqrt{(\xi_1^2+\xi_2^2)/2}\gg 1$. Presently available strong lasers have photon energies of the order of $1\;\text{eV}$ and deliver intensities of the order of $10^{22}\;\text{W/cm$^2$}$, which correspond to values of $\xi$ of the order of $100$ \cite{Emax}. If $b\ll 1$ and $\xi\gg 1$ such that $b\xi$ is fixed (which corresponds to the quasiclassical limit), the total rate $\dot{W}_1$ depends on the strong field only through the two gauge- and Lorentz-invariant parameters
\begin{equation}
\chi_{1,2}= b\xi_{1,2}=\frac{E_{1,2}}{E_{cr}}\frac{m}{\omega_1}.
\end{equation}
After the expansion over $b$, the functions introduced above simplify to
\begin{eqnarray}
\label{F_1_2}
&&F_1=\rho^2\chi^2f, \qquad F_2=\frac{4w^2\chi^2}{\mathcal{Z}_0^2}\left(\frac{\sin\rho}{\rho}-\cos\rho\right)^2g,\nonumber\\
\label{F_3_Z_0}
&&F_3=\frac{4w^2\chi^2}{\mathcal{Z}_0}\rho\sin\rho\left(\frac{\sin\rho}{\rho}-\cos\rho\right)g, \qquad \mathcal{Z}_0=1+\frac{\chi^2\rho^2}{3}f,\nonumber\\
\label{f_g}
&&f=1-\mu_s\cos\psi, \qquad g=1+\mu_w\mu_s-(\mu_s+\mu_w)\cos\psi,
\end{eqnarray}
where we have used the following notation:
\begin{eqnarray}
&&\chi=\sqrt{\frac{\chi_1^2+\chi_2^2}{2}}, \qquad \mu_s=\frac{\chi_1^2-\chi_2^2}{\chi_1^2+\chi_2^2}, \nonumber\\
\label{w}
&& w=\sqrt{\frac{w_1^2+w_2^2}{2}}, \qquad \mu_w=\frac{w_1^2-w_2^2}{w_1^2+w_2^2}.
\end{eqnarray}
The parameters $\mu_s$ and $\mu_w$ describe the ellipticities of the strong and of the weak field, respectively. Note that $\dot{W}_1$ depends linearly on the polarization of the high-energy photon through the function $g$ while the dependence of $\dot{W}_1$ on $\mu_s$ is complicated. 

Let us discuss why at $\xi\gg 1$ the correction to the Bethe-Heitler process depends on the parameter $\chi$ and not on $\xi$ itself. At $\xi\gg 1$ the effect of the laser field on a charged particle can be treated in the quasiclassical approximation. Then, the function $\mathbf{\Delta}(y)$ for the strong field in Eq. (\ref{Not_1}), that is always present in the expression of $\dot{W}_1$ (see for example Eq. (\ref{W_dot_i})), has the following meaning. It is the ratio $\delta \bm{p}_{\perp}/m$, with $\delta \bm{p}_{\perp}$ being the momentum transfer, perpendicular to $\hat{\mathbf{z}}$, from the laser field to the electron and positron during the formation time $t_0$ of the pure Bethe-Heitler process (without any laser field). The physical meaning of the quantity $\rho$ is the phase shift of the weak high-energy field during the time $t_0$ while $b\rho$ is the corresponding phase shift of the strong low-frequency field. Since the main contribution to the process comes from the region of integration $\rho\lesssim 1$ (corresponding to longitudinal momentum transfer in units of the electron mass: $Qx\lesssim m/\omega_1\ll 1$) and $b=\omega_0/\omega_1\ll 1$, the phase shift of the strong field is much smaller than unity, we can expand the function $\mathbf{\Delta}(y)$ for the strong laser field on this parameter and obtain that $|\mathbf{\Delta}(y)|\propto \chi$. Another intuitive way to understand why the parameter $\chi$ controls the influence of the laser field on the process at hand is that $\chi$ is the ratio between the energy $\Delta \mathcal{E}_L$ transferred by the laser to the pair in the process, i. e. $\Delta \mathcal{E}_L\sim eE/m$ and the energy $\mathcal{E}_{\gamma}$ that the electron-positron pair would have without laser field, i. e. $\mathcal{E}_{\gamma}\sim\omega_1$. Finally, one can also interpret the parameter $\chi$ as the transverse momentum transfer by the laser field in one wavelength of the weak field in units of the electron mass.
%
%
\subsection{Asymptotics of the rate and numerical results}

The rate $\dot{W}_1$ in the quasiclassical limit given by Eq. (\ref{W_dot}) with the substitutions in Eq. (\ref{f_g}) at small $\chi$ has the form $\dot{W}_1=\dot{W}_{BH}+\delta\dot{W}_1$ where
\begin{equation}
\label{W_dot_BH}
\dot{W}_{BH}=\frac{7}{9}\frac{(Z\alpha)^2\omega_1}{\pi}w^2\left[\ln\left(\frac{2\omega_1}{m}\right)-\frac{109}{42}\right]
\end{equation}
corresponds to the pure Bethe-Heitler process for $\omega_1\gg m$ and the leading correction $\delta\dot{W}_1$ is given by
\begin{equation}
\label{delta_dot_weak}
\delta\dot{W}_1=-\frac{19(Z\alpha)^2}{12\sqrt{3}}\frac{\omega_1\chi w^2}{\pi} \int_0^{2\pi}\frac{d\psi}{2\pi}\sqrt{f}\left[\ln\left(\frac{\omega_1\chi \sqrt{f}}{2m\sqrt{3}}\right)-C+\frac{11}{57}\right].
\end{equation}
The cross section in Eq. (\ref{eq:BHformula}) is obtained by dividing $\dot{W}_{BH}$ by the incoming photon flux $j=\omega_1m^2w^2/4\pi\alpha$. Note that the expression of the above correction which is negative, is valid if the argument of the logarithm ($\sim \chi\omega_1/m$) is much larger than unity which is equivalent to $E/E_{cr}\gg 1$. In contrast to the Bethe-Heitler rate which is independent of both $\mu_s$ and $\mu_w$, the correction $\delta\dot{W}_1$ depends on the polarization of the strong laser field.

In the opposite limit $\chi\gg 1$ of strong fields the asymptotic of $\dot{W}_1$ reads:
\begin{equation}
\label{W_dot_l}
\dot{W}_1= \frac{13(Z\alpha)^2}{24\sqrt{3}}\frac{\omega_1w^2}{\pi\chi}\int_0^{2\pi}\frac{d\psi}{2\pi}\frac{1}{\sqrt{f}}\left\{\left(2-\frac{g}{f}\right)\left[\ln\left(\frac{\omega_1\chi \sqrt{f}}{2m\sqrt{3}}\right)-C\right]-\frac{38}{39}+\frac{58}{39}\frac{g}{f}\right\},
\end{equation}
where the functions $f$ and $g$ are defined in Eq. (\ref{f_g}). Since the asymptotic (\ref{W_dot_l}) is proportional to $\chi^{-1}$, it is much smaller that $\dot{W}_{BH}$.

Let us discuss now the influence of the laser field at intermediate values of the parameter $\chi$. The dependence of the ratio $\dot{W}_1/\dot{W}_{BH}$ on $\chi$ is shown in Fig. 1 for $\omega_1=100\,m$. The following paradigmatic situations are considered: strong laser field circularly polarized and high-energy photon arbitrarily polarized (continuous curve), strong laser field linearly polarized and high-energy photon circularly polarized (dashed curve), strong laser field linearly polarized and high-energy photon linearly polarized in the same direction as the strong laser field (dotted curve), strong laser field linearly polarized and high-energy photon linearly polarized in the perpendicular direction with respect to the strong laser field (dashed-dotted curve). In all cases we observe a suppression of the pair-production rate due to the effect of the laser field. This suppression induced by the laser field is typical also for other external electromagnetic field configurations like, for instance, a constant magnetic field \cite{Baier_b_1994,Baier_2005} (note that in the limit $\xi\gg 1$ and for undercritical magnetic fields similar effects in the two cases are expected from general considerations \cite{Ritus_Review}). In fact, at very high energies of the incoming photon ($\omega_1\gg m$) the formation region  (coherence length) of the process becomes relatively large: in our problem it is of the order of $\omega_1/m^2\gg \lambda_c$ along the propagation direction of the incoming photon. The presence of the external field perturbs the system by deviating the electron and the positron along the transverse direction, effectively reducing the formation region and then the total rate of the process. We note that our process is connected by crossing symmetry to the bremsstrahlung process in a laser field. An analogous suppression in the cross section has been predicted in bremsstrahlung in the presence of a magnetic field \cite{Baier_b_1994,Baier_2005}.

It is interesting that the amount of suppression depends on the mutual polarization of the strong laser field and of the high-energy photon. The largest suppression is observed in the case in which the strong laser field and the high-energy photon are both linearly polarized along the same direction. Also, it can be seen from the general expression Eq. (\ref{W_dot}) (see also Eq. (\ref{f_g})) that if the strong laser field is circularly polarized, then the rate $\dot{W}_1$ is independent of the high-energy photon polarization. Finally, we observe that already at small values of the parameter $\chi$ we have a relatively strong suppression of the pair-production rate (about $10\;\%$ at $\chi=0.2$).

In order to emphasize the role of the mutual polarization of the strong laser field and of the high-energy photon we present in Fig. 2 and in Fig. 3 the ratio  $\dot{W}_1/\dot{W}_{BH}$ as a function of $\mu_w$ (Fig. 2) and of $\mu_s$ (Fig. 3). In both cases we have set $\chi=0.2$ and $\omega_1=100\,m$. In Fig. 2 the strong field is linearly polarized, the dependence of $\dot{W}_1/\dot{W}_{BH}$ on $\mu_w$ is a straight line and a difference of about $5\;\%$ is observed between the two extreme cases $\mu_w=-1$ (polarization perpendicular to that of the strong laser field) and $\mu_w=+1$ (polarization parallel to that of the strong laser field). In Fig. 3 the two different situations in which the high-energy photon is circularly polarized (continuous curve) and linearly polarized (dashed line) are plotted. 

In Figs. 2 and 3 the value of $\chi$ is small. However, it is impossible to explain the features of these figures from our leading correction $\delta\dot{W}_1$ in Eq. (\ref{delta_dot_weak}) and it is necessary to calculate the next-to-leading correction with respect to $\chi$. As we pointed out above, the linear dependence of $\dot{W}_1$ on $\mu_w$ in the semiclassical limit holds for any values of the parameter $\chi$. The next-to-leading correction proportional to $\mu_w$ and calculated in the logarithmic approximation has the form
\begin{equation}
\delta\dot{W}'_1=-\frac{7(Z\alpha)^2}{9}\frac{\omega_1\chi^2 w^2\mu_s\mu_w}{\pi}\left[\ln\left(\frac{2\omega_1}{m}\right)-\frac{257}{84}\right].
\end{equation}
The value of the slope of the straight line in Fig. 2 agrees well with the prediction from the above equation. Moreover, the proportionality of $\delta\dot{W}'_1$ to $-\mu_w$ explains the symmetry of the continuous line ($\mu_w=0$) and the asymmetry of the dashed line ($\mu_w=+1$) in Fig. 3.

In order to give the impression of the size of the effect discussed above, we used the value of the intensity of the strong laser field $I=5\times 10^{23}\;\text{W/cm$^2$}$, $\omega_1$ in the laboratory frame $\omega_{1,\text{lab}}=3.7\;\text{keV}$ and the proton ($Z=1$) Lorentz factor $\gamma=7000$ (which corresponds to the proton energy available at the LHC). In this case we obtain $\chi=0.2$ and a suppression of about $10\;\%$ (note that the contribution of the constant terms to the quantity $\delta\dot{W}_1$ in Eq. (\ref{delta_dot_weak}) is about $20\,\%$). Laser intensities of this order of magnitude can be available with the next generation of Petawatt laser systems, making possible in principle the measurement of the high-field QED effect discussed here. It is also important to mention that some of these Petawatt lasers would have a relative small size (see, for example, the homepage \cite{PFS} of the table-top Petawatt laser system under construction in Garching (Germany)). Since strong laser fields like that considered in the above example are obtained experimentally by tightly focusing the field, we want to estimate here qualitatively the alteration induced into the above results by the tight focusing of the laser field. We first observe that the parameter $\chi$ can be written as $\chi=(k_0u)\xi/(k_1u)$, where $k_0^{\mu}=(\omega_0,\mathbf{k}_0)$ ($k_1^{\mu}=(\omega_1,\mathbf{k}_1)$) is the four-wavevector of the strong (weak) field and $u=\gamma(1,\mathbf{v})$ is the relativistic four-velocity of the nucleus. Now, if $w_0$ is the waist size of the strong field, the laser photons propagate within a cone with an aperture $\vartheta$ of the order of $w_0/l_R$, with $l_R=\omega_0w_0^2/2$ being the laser's Rayleigh length. By indicating as $\chi_{\vartheta}$ the parameter $\chi$ for those photons propagating at an angle $\vartheta\ll 1$ with respect to the vector $-\mathbf{v}$, we have $\chi_{\vartheta}\approx\chi(1-\vartheta^2/4)$ at $\vartheta\ll 1$. In this sense, even if the strong field is focused up to one wavelength the induced correction is about $2.5\;\%$ and it is smaller than the predicted suppression induced by the strong laser field on the total rate.
Finally, we point out that in the above example, many pairs are expected to be created by the strong laser field and the proton alone. However, the background generated by those pairs can be in principle determined experimentally by performing the experiment twice: the first time without and the second time with the high-energy photon beam. As a concluding remark, one has also to note that in view of an experimental realization of the discussed effect, a deeper analysis should be performed about other possible background processes (cascade pair production processes, for example) and also modifications of the pairs kinematics due to radiation back reaction. However, at the current stage these investigations are beyond the present paper.
%
%
\section{Near-threshold photoproduction}

In this Section we consider the interesting case $0<2m-\omega_1\ll m$, $\xi\gg 1$ and $\omega_0\ll m$ (quasiclassical limit) and $\chi\ll 1$. This case has been already reported without technical details in \cite{Di_Piazza_2009}. The physical scenario here is very different from that considered above. In fact, under the present conditions there is no Bethe-Heitler process without laser field and the pair creation occurs through tunneling. On the other hand, the pair production by combined laser and Coulomb fields is exponentially suppressed by a factor $\exp(-\sqrt{3}/\chi)$ \cite{Yakovlev_1966}. The present parameter regime is interesting as one can expect that the small energy gap between the photon energy $\omega_1$ and the threshold $2m$ should essentially amplify the electron-positron rate as compared to the case of combined laser and Coulomb field. As we will see, this allows the possibility of observing tunneling pair creation at strong electric field amplitudes (in the laboratory frame) much smaller than $E_{cr}$ (see also \cite{Di_Piazza_2009}).

Starting from the general expression in Eq. (\ref{W_dot_i}) of the photoproduction rate and expanding it with respect to the amplitude $w$ of the weak field up to quadratic terms, we arrive at
\begin{equation}
\label{W_dot_t}
\begin{split}
\dot{W}&=-\frac{(Z\alpha)^2mw^2}{\pi^2}\text{Im}\int_0^{\infty}\frac{dQ}{Q^2}\int_0^1dv\int_0^{\infty}\frac{d\rho}{\rho}\int_0^{2\pi}\frac{d\psi}{2\pi}\int_0^1dx\frac{1-x^2}{x^2}e^{-i\Phi}\\
&\times\left\{i\frac{a\rho}{Qx(1-v^2)}\left[1-2\rho^2\chi^2f\frac{1+v^2}{1-v^2}+Q^2\frac{1-3v^2+x^2(1+v^2)}{1-x^2}\right]\right.\\
&\times\left[1-\frac{\sin^2\rho}{\rho^2}-i\frac{a\rho\chi^2}{Qx(1-v^2)}g\left(\frac{\sin\rho}{\rho}-\cos\rho\right)^2\right]\\
&\left.-\frac{1+v^2}{1-v^2}\sin\rho\left[i\frac{4a\rho^2\chi^2}{Qx(1-v^2)}\left(\frac{\sin\rho}{\rho}-\cos\rho\right)g-2\sin\rho\right]\right\},\\
\Phi&=\frac{a\rho[\mathcal{Z}_0+Q^2(1-v^2)]}{Qx(1-v^2)},
\end{split}
\end{equation}
where $a=2m/\omega_1>1$ and $\mathcal{Z}_0$, $f$ and $g$ are given in Eq. (\ref{F_3_Z_0}). The phase $\Phi$ is always positive and larger than $2a\rho$. At $\chi\ll 1$, the main contribution to the rate $\dot{W}$ in Eq. (\ref{W_dot_t}) arises from the region of integration $\rho\gg 1$ under compensation between the exponential $\exp(-i\Phi)$ and $\exp(2i\rho)$ coming from the trigonometric functions in the pre-exponent. Besides, this compensation requires that $1-x\ll 1$ and $v\ll 1$. Taking the integral over $x$ and $v$ in this region we obtain
\begin{equation}
\label{W_dot_threshold}
\begin{split}
\dot{W}&=\frac{(Z\alpha)^2mw^2}{\pi^{3/2}}\text{Im}\int_0^{2\pi}\frac{d\psi}{2\pi}\int_0^{\infty}dQ\sqrt{Q}\int_0^{\infty}\frac{d\rho}{\rho^{7/2}}\exp\left\{i\rho\left[2-\frac{a}{Q}(\mathcal{Z}_0+Q^2)\right]-i\frac{\pi}{4}\right\}\\
&\times\frac{1}{\sqrt{\mathcal{Z}_0}}\frac{1}{(\mathcal{Z}_0+Q^2)^2}\left\{\frac{i}{4\rho Q}\left[1+i\rho Q(\mathcal{Z}_0+Q^2)-2\rho^2\chi^2f\right]\left(1-i\frac{\rho^3\chi^2}{Q}g\right)+\frac{\rho^2\chi^2}{Q}g-\frac{1}{2}\right\}.
\end{split}
\end{equation}
Then, we take the integral first over $Q$ and then over $\rho$ by means of the saddle point method. The saddle point in the integration over $Q$ is $Q_0=\sqrt{\mathcal{Z}_0}$ and that over $\rho$ is $\rho_0=-i\sqrt{\delta/f\chi^2}$ with $\delta=a^2-1\ll 1$. Finally, we divide $\dot{W}$ by the flux $j=\omega_1m^2w^2/4\pi\alpha\approx m^3w^2/2\pi\alpha$ and arrive at the cross section
\begin{equation}
\label{W_dot_tf}
\sigma=\frac{\sqrt{\pi}\alpha(Z\alpha)^2\chi^2}{8m^2}\sqrt{\zeta}\int_0^{2\pi}\frac{d\psi}{2\pi}f^{1/4}\left(g+2\zeta f^{3/2}\right)\exp\left(-\frac{2}{3\zeta\sqrt{f}}\right),
\end{equation}
which is valid if $\zeta=\chi/\delta^{3/2}\ll 1$. Formally the last term in the pre-exponent is much smaller than the first one. However, if the strong laser field and the high-energy photon are linearly polarized in perpendicular directions, we have $g=0$ and the non-zero contribution is given by the second term. As expected, in our case the suppression due to the exponential factor is much smaller than that in the case of combined laser and Coulomb field.

The above expression (\ref{W_dot_tf}) is valid for any polarization of the incoming high-energy photon and of the strong laser field. In particular, if the strong laser field is circularly polarized, we obtain:
\begin{equation}
\label{sigma_c}
\sigma=\frac{\sqrt{\pi}\alpha(Z\alpha)^2\chi^2}{8m^2}\sqrt{\zeta}\exp\left(-\frac{2}{3\zeta}\right).
\end{equation}
If $\mu_s$ is not too small, i. e. if $|\mu_s|\gg \zeta$ we have
\begin{equation}
\label{sigma_l}
\sigma=\frac{\sqrt{3}\alpha(Z\alpha)^2\chi^2}{8\sqrt{2}m^2}\zeta\frac{\kappa^2}{\sqrt{\kappa-1}}\left[1+\text{sgn}(\mu_s)\mu_w+2\zeta\sqrt{\kappa}\right]\exp\left(-\frac{2}{3\zeta\sqrt{\kappa}}\right),
\end{equation}
with $\kappa=1+|\mu_s|$.

It is of interest to show qualitatively how the exponential behaviour in Eqs. (\ref{sigma_c}) and (\ref{sigma_l}) arises. We consider the paradigmatic situation of an electron in the negative continuum that absorbs a photon with frequency $\omega_1$ and then, due to a constant and uniform electric field $E$, it tunnels to the positive continuum. The width $l$ of the barrier the electron has to tunnel is approximately given by $eEl=2m-\omega_1\approx m\delta\ll m$. Therefore, in the quasiclassical limit one obtains
\begin{equation}
\sigma=\dot W/j\sim\frac{1}{m^2}\exp\left[-2\int_0^ldx\sqrt{2m(m\delta-eEx)}\right]=\frac{1}{m^2}\exp\left(-\frac{2\sqrt{2}}{3\zeta}\right),
\end{equation}
which qualitatively reproduces the exponential dependence in Eqs. (\ref{sigma_c}) and (\ref{sigma_l}).

In \cite{Di_Piazza_2009} we had shown the dependence of the production rate on the parameter $\zeta$ for two fixed mutual polarizations of the strong and the weak field (see Fig. 3 in \cite{Di_Piazza_2009}). In Fig. 4 we plot here the ratio $\sigma/\sigma_0$ with $\sigma_0=\alpha(Z\alpha)^2\chi^2/m^2$ and $\sigma$ given by Eq. (\ref{W_dot_tf}) for a fixed value of $\zeta=0.1$ and of $\mu_w$ as a function of the ellipticity $\mu_s$ of the strong laser field. The continuous (dashed) curve shows the case of a circularly (linearly) polarized weak laser field with $\mu_w=0$ ($\mu_w=+1$). It is seen that the effect of the strong laser field is much more pronounced in the case of linear polarization when it is polarized along the same direction as the weak field ($\mu_s=+1$). 

As an example, we consider a proton ($Z=1$) with an energy of $980\;\text{GeV}$ (proton beams with such an energy are already available at Tevatron \cite{PDG}) and a laser field with an intensity $I=5\times 10^{20}\;\text{W/cm$^2$}$, a wavelength of $0.8\;\text{$\mu$m}$, a spot radius of $10\;\text{$\mu$m}$, a pulse duration of $5\;\text{fs}$ and a repetition rate of $10\;\text{Hz}$ (Petawatt lasers with a such high repetition rate have been proposed, see e. g. \cite{PFS}). The photon energy in the laboratory frame corresponding to the threshold $2m$ in the rest frame is $\omega_{th}=2m/2\gamma=490\;\text{eV}$ and we set $\Delta\omega_1/\omega_1=(\omega_{th}-\omega_1)/\omega_1=0.05$. With the above values we have $\chi=0.05$ and $\delta=0.1$ and by considering the most favorable case $\mu_s=\mu_w=+1$, the total cross section is $\sigma=0.5\;\text{$\mu$barn}$. Note that since $\zeta=1.5$, the analytical asymptotic in Eq. (\ref{sigma_l}) slightly overestimates the cross section and the above value of $\sigma$ has been obtained numerically starting directly from Eq. (\ref{W_dot_threshold}). As a source of the X-ray photons we consider the table-top X-FEL proposed in \cite{Gruener_2007} in which the electron beam is generated by a laser-plasma accelerator ($8\times 10^{11}$ photons per bunch, a pulse duration of $4\;\text{fs}$ and a bandwidth of $0.2\;\%$). If this beam is focused to a radius of $10\;\text{$\mu$m}$, one obtains a photon flux of $6.4\times 10^{31}\;\text{s$^{-1}$cm$^{-2}$}$ and about $1.3\times 10^{-13}$ pairs per laser shot and per proton. Finally, by considering the values of the proton beams at the Tevatron, i. e. $2.4\times 10^{11}$ protons per bunch, bunch length of $50\;\text{cm}$ and beam size of $29\;\text{$\mu$m}$ \cite{PDG}, we obtain about one pair every five hours. We note that in the case of combined Coulomb and laser field alone, the number of pairs obtained is completely negligible with the above physical parameters. Other numerical examples have been already presented in \cite{Di_Piazza_2009} showing the possibility of observing the discussed effect with presently available laser and accelerator technology.

\section{Conclusion}
In this paper we have calculated the total electron-positron pair production rate in the collision of a high-energy photon, a strong laser field and a relativistic nucleus. The strong laser field has been taken into account exactly in the calculations while the high-energy photon and the nuclear field only in the leading order. We have considered two situations. In the first one the frequency of the high-energy photon in the rest frame of the nucleus is much larger than the electron mass. In this case we have found that the laser field may significantly suppress the cross section of the Bethe-Heitler process at soon available values of parameters. In the second case the frequency of the high-energy photon in the rest frame of the nucleus is close to and smaller than the pair production threshold. In this case pair production does not occur without the strong laser field. The effect of the laser field leads to a relatively large cross section of pair production also with proton beam parameters and laser technology already available. In both cases the mutual polarization of the strong laser field and of the high-energy photon is very important for the value of the electron-positron production rate.

%
%

\section*{Acknowledgments}
E. L. is grateful for the kind hospitality shown to him during a short visit at the Budker Institute of Nuclear Physics. A. I. M. gratefully acknowledges the hospitality and the financial support he has received during his visit at Max-Planck-Institute for Nuclear Physics. The work was supported in part by the RFBR under the grants 09-02-00024 and 08-02-91969, and in part by the JSPS.

\clearpage

\begin{figure}
\begin{center}
\includegraphics[width=10cm]{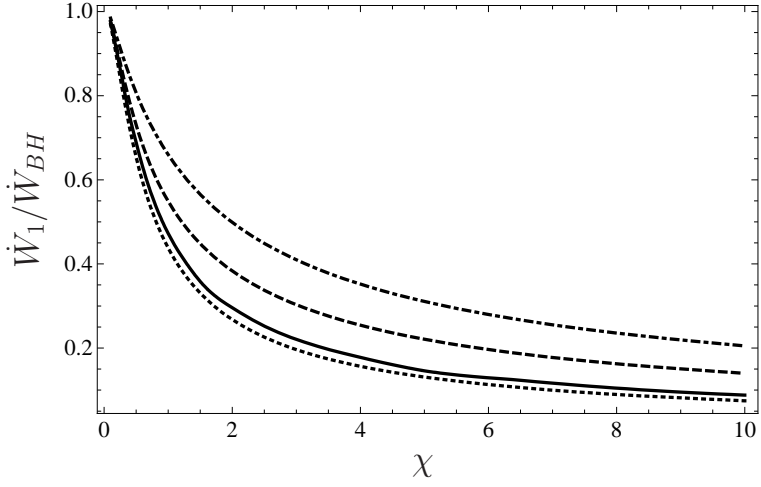}
\end{center}
\caption{The ratio $\dot{W}_1/\dot{W}_{BH}$ as a function of the parameter $\chi$. The energy of the high-energy photon is $\omega_1=100\,m=51\;\text{MeV}$ and the following paradigmatic situations are shown: strong laser field circularly polarized and high-energy photon arbitrarily polarized (continuous curve), strong laser field linearly polarized and high-energy photon circularly polarized (dashed curve), strong laser field linearly polarized and high-energy photon linearly polarized in the same direction as the strong laser field (dotted curve), strong laser linearly polarized and high-energy photon linearly polarized in the perpendicular direction with respect to the strong laser field (dashed-dotted curve).}
\end{figure}

\clearpage

\begin{figure}
\begin{center}
\includegraphics[width=10cm]{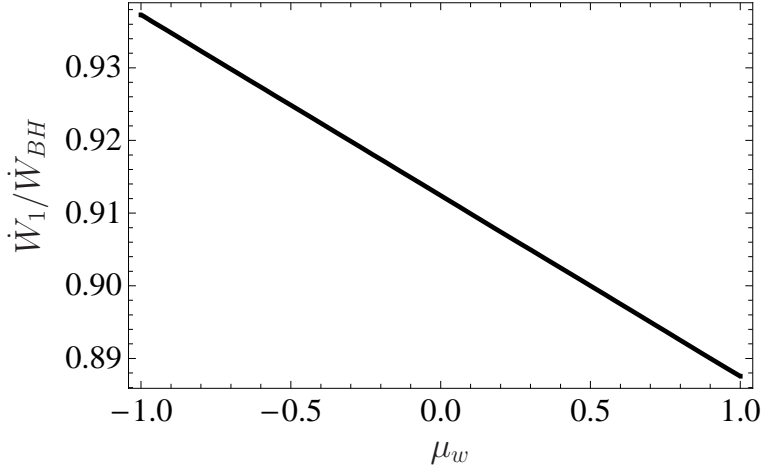}
\end{center}
\caption{The ratio $\dot{W}_1/\dot{W}_{BH}$ as a function of the ellipticity of the high-energy photon $\mu_w$. The parameters $\omega_1$ and $\chi$ have been set to $100\,m$ and to $0.2$, respectively. The strong laser field is linearly polarized and $\mu_w=-1$ ($\mu_w=+1$) corresponds to the situation in which the high-energy photon is linearly polarized in the perpendicular direction of (same direction as) the strong laser field.}
\end{figure}

\clearpage

\begin{figure}
\begin{center}
\includegraphics[width=10cm]{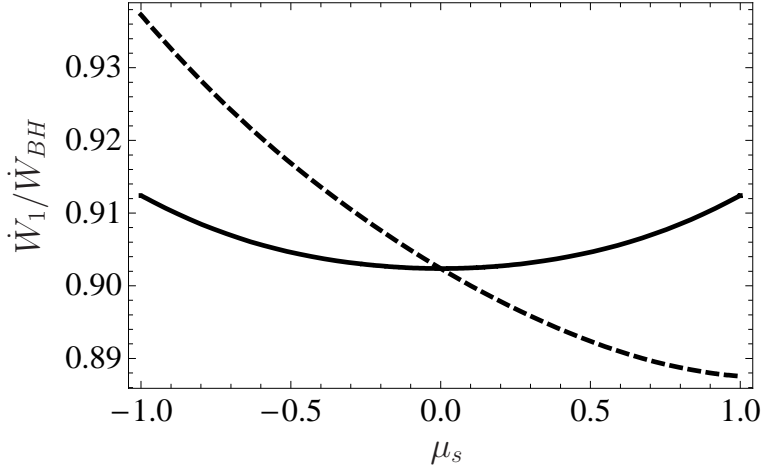}
\end{center}
\caption{The ratio $\dot{W}_1/\dot{W}_{BH}$ as a function of the ellipticity of the strong laser field $\mu_s$. The parameters $\omega_1$ and $\chi$ have been set to $100\,m$ and to $0.2$, respectively. The continuous (dashed) curve refers to the case in which the high-energy photon is circularly $\mu_w=0$ (linearly $\mu_w=+1$) polarized, respectively. The value $\mu_s=-1$ ($\mu_s=+1$) corresponds to the situation in which the strong laser field is linearly polarized in the perpendicular direction of (same direction as) the high-energy photon.}
\end{figure}

\clearpage

\begin{figure}
\begin{center}
\includegraphics[width=10cm]{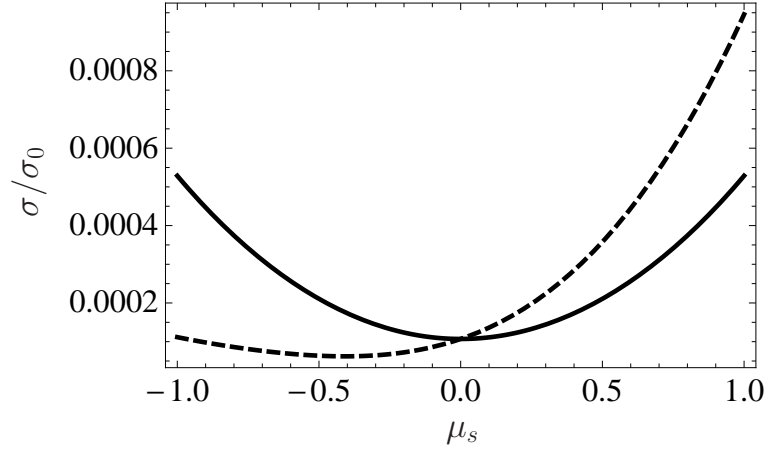}
\end{center}
\caption{The ratio $\sigma/\sigma_0$ with $\sigma_0=\alpha(Z\alpha)^2\chi^2/m^2$ and $\sigma$ given by Eq. (\ref{W_dot_tf}) at $\zeta=0.1$ and at $\mu_w=0$ (circularly polarized weak laser field; continuous curve) and at $\mu_w=+1$ (linearly polarized weak laser field; dashed curve).}
\end{figure}


\clearpage

\begin{thebibliography}{39}
\expandafter\ifx\csname natexlab\endcsname\relax\def\natexlab#1{#1}\fi
\expandafter\ifx\csname bibnamefont\endcsname\relax
  \def\bibnamefont#1{#1}\fi
\expandafter\ifx\csname bibfnamefont\endcsname\relax
  \def\bibfnamefont#1{#1}\fi
\expandafter\ifx\csname citenamefont\endcsname\relax
  \def\citenamefont#1{#1}\fi
\expandafter\ifx\csname url\endcsname\relax
  \def\url#1{\texttt{#1}}\fi
\expandafter\ifx\csname urlprefix\endcsname\relax\def\urlprefix{URL }\fi
\providecommand{\bibinfo}[2]{#2}
\providecommand{\eprint}[2][]{\url{#2}}

\bibitem[{\citenamefont{Bethe and Heitler}(1934)}]{BeHe1934}
\bibinfo{author}{\bibfnamefont{H.~A.} \bibnamefont{Bethe}} \bibnamefont{and}
  \bibinfo{author}{\bibfnamefont{W.}~\bibnamefont{Heitler}},
  \bibinfo{journal}{Proc. Roy. Soc. A} \textbf{\bibinfo{volume}{146}},
  \bibinfo{pages}{83} (\bibinfo{year}{1934}).

\bibitem[{\citenamefont{Milstein and Schumacher}(1994)}]{Milstein1994183}
\bibinfo{author}{\bibfnamefont{A.~I.} \bibnamefont{Milstein}} \bibnamefont{and}
  \bibinfo{author}{\bibfnamefont{M.}~\bibnamefont{Schumacher}},
  \bibinfo{journal}{Phys. Rep.} \textbf{\bibinfo{volume}{243}},
  \bibinfo{pages}{183 } (\bibinfo{year}{1994}).

\bibitem[{\citenamefont{Akhmadaliev et~al.}(1998)\citenamefont{Akhmadaliev,
  Kezerashvili, Klimenko, Malyshev, Maslennikov, Milov, Milstein, Muchnoi,
  Naumenkov, Panin et~al.}}]{AkhKezetal1998}
\bibinfo{author}{\bibfnamefont{{\relax Sh}.~{\relax Zh}.}
  \bibnamefont{Akhmadaliev}} \bibnamefont{et~al.}, \bibinfo{journal}{Phys. Rev. C}
  \textbf{\bibinfo{volume}{58}}, \bibinfo{pages}{2844} (\bibinfo{year}{1998}).

\bibitem[{\citenamefont{Lee et~al.}(2003)\citenamefont{Lee, Maslennikov,
  Milstein, Strakhovenko, and Tikhonov}}]{Leeetal2003}
\bibinfo{author}{\bibfnamefont{R.~N.} \bibnamefont{Lee}} \bibnamefont{et~al.}, \bibinfo{journal}{Phys. Rep.}
  \textbf{\bibinfo{volume}{373}}, \bibinfo{pages}{213} (\bibinfo{year}{2003}).

\bibitem[{\citenamefont{Akhmadaliev et~al.}(2002)\citenamefont{Akhmadaliev,
  Kezerashvili, Klimenko, Lee, Malyshev, Maslennikov, Milov, Milstein, Muchnoi,
  Naumenkov et~al.}}]{AkhKezetal2002}
\bibinfo{author}{\bibfnamefont{{\relax Sh}.~{\relax Zh}.}
  \bibnamefont{Akhmadaliev}} \bibnamefont{et~al.}, \bibinfo{journal}{Phys. Rev.
  Lett.} \textbf{\bibinfo{volume}{89}}, \bibinfo{pages}{061802}
  (\bibinfo{year}{2002}).

\bibitem[{\citenamefont{Bethe and Maximon}(1954)}]{BethMax1954}
\bibinfo{author}{\bibfnamefont{H.~A.} \bibnamefont{Bethe}} \bibnamefont{and}
  \bibinfo{author}{\bibfnamefont{L.~C.} \bibnamefont{Maximon}},
  \bibinfo{journal}{Phys. Rev.} \textbf{\bibinfo{volume}{93}},
  \bibinfo{pages}{768} (\bibinfo{year}{1954}).

\bibitem[{\citenamefont{\O{}verb\o{} et~al.}(1973)\citenamefont{\O{}verb\o{},
  Mork, and Olsen}}]{OvMoOl1973}
\bibinfo{author}{\bibfnamefont{I.}~\bibnamefont{\O{}verb\o{}}},
  \bibinfo{author}{\bibfnamefont{K.~J.} \bibnamefont{Mork}}, \bibnamefont{and}
  \bibinfo{author}{\bibfnamefont{H.~A.} \bibnamefont{Olsen}},
  \bibinfo{journal}{Phys. Rev. A} \textbf{\bibinfo{volume}{8}},
  \bibinfo{pages}{668} (\bibinfo{year}{1973}).

\bibitem[{\citenamefont{Lee et~al.}(2004)\citenamefont{Lee, Milstein, and
  Strakhovenko}}]{LeeMilStr2004}
\bibinfo{author}{\bibfnamefont{R.~N.} \bibnamefont{Lee}},
  \bibinfo{author}{\bibfnamefont{A.~I.} \bibnamefont{Milstein}},
  \bibnamefont{and} \bibinfo{author}{\bibfnamefont{V.~M.}
  \bibnamefont{Strakhovenko}}, \bibinfo{journal}{Phys. Rev. A}
  \textbf{\bibinfo{volume}{69}}, \bibinfo{pages}{022708}
  (\bibinfo{year}{2004}).

\bibitem[{\citenamefont{Lee et~al.}(2005)\citenamefont{Lee, Milstein,
  Strakhovenko, and Schwarz}}]{LeeMilStr2005}
\bibinfo{author}{\bibfnamefont{R.~N.} \bibnamefont{Lee}} \bibnamefont{et~al.}, \bibinfo{journal}{Zh. \'{E}ksp. Teor. Fiz.}
  \textbf{\bibinfo{volume}{127}}, \bibinfo{pages}{5} (\bibinfo{year}{2005}),
  \bibinfo{note}{[JETP \textbf{100}, 1 (2005)]}.

\bibitem[{\citenamefont{Baur et~al.}(2007)\citenamefont{Baur, Hencken, and
  Trautmann}}]{BaHeTra2007}
\bibinfo{author}{\bibfnamefont{G.}~\bibnamefont{Baur}} \bibnamefont{et~al.},
  \bibinfo{journal}{Phys. Rep.} \textbf{\bibinfo{volume}{453}},
  \bibinfo{pages}{1} (\bibinfo{year}{2007}).

\bibitem[{\citenamefont{Baltz et~al.}(2008)\citenamefont{Baltz, Baur,
  d'Enterria, Frankfurt, Gelis, Guzey, Hencken, Kharlov, Klasen, Klein
  et~al.}}]{Baltz20081}
\bibinfo{author}{\bibfnamefont{A.~J.} \bibnamefont{Baltz}} \bibnamefont{et~al.}, \bibinfo{journal}{Phys. Rep.}
  \textbf{\bibinfo{volume}{458}}, \bibinfo{pages}{1 } (\bibinfo{year}{2008}).

\bibitem[{\citenamefont{Reiss}(1962)}]{Re1962}
\bibinfo{author}{\bibfnamefont{H.~R.} \bibnamefont{Reiss}},
  \bibinfo{journal}{J. Math. Phys.} \textbf{\bibinfo{volume}{3}},
  \bibinfo{pages}{59} (\bibinfo{year}{1962}).

\bibitem[{\citenamefont{Nikishov and Ritus}(1964)}]{NiRi1964}
\bibinfo{author}{\bibfnamefont{A.~I.} \bibnamefont{Nikishov}} \bibnamefont{and}
  \bibinfo{author}{\bibfnamefont{V.~I.} \bibnamefont{Ritus}},
  \bibinfo{journal}{Zh. \'{E}ksp. Teor. Fiz.} \textbf{\bibinfo{volume}{46}},
  \bibinfo{pages}{776} (\bibinfo{year}{1964}), \bibinfo{note}{[Sov. Phys. JETP
  \textbf{19}, 529 (1964)]}.



\bibitem{Mueller_2003} C. M\"{u}ller, A. B. Voitkiv and N. Gr\"{u}n, Phys. Rev. Lett. \textbf{91}, 223601 (2003); 
\textit{ibid.}, Phys Rev. A \textbf{67}, 063407 (2003); \textit{ibid.}, Phys Rev. A \textbf{70}, 023412 (2004).

\bibitem[{\citenamefont{Milstein et~al.}(2006)\citenamefont{Milstein, M\"uller,
  Hatsagortsyan, Jentschura, and Keitel}}]{MiMuHaJeKe2006}
\bibinfo{author}{\bibfnamefont{A.~I.} \bibnamefont{Milstein}} \bibnamefont{et~al.}, \bibinfo{journal}{Phys. Rev. A}
  \textbf{\bibinfo{volume}{73}}, \bibinfo{pages}{062106}
  (\bibinfo{year}{2006}).

\bibitem[{\citenamefont{Kuchiev and Robinson}(2007)}]{KuRo2007} M. Yu. Kuchiev, Phys. Rev. Lett. \textbf{99}, 130404 (2007);
\bibinfo{author}{\bibfnamefont{M.~{\relax Yu}.} \bibnamefont{Kuchiev}}
  \bibnamefont{and} \bibinfo{author}{\bibfnamefont{D.~J.}
  \bibnamefont{Robinson}}, \bibinfo{journal}{Phys. Rev. A}
  \textbf{\bibinfo{volume}{76}}, \bibinfo{pages}{012107}
  (\bibinfo{year}{2007}).

\bibitem{Deneke_2008} C. M\"{u}ller, C. Deneke and C. H. Keitel, Phys. Rev. Lett. \textbf{101}, 060402 (2008);
C. M\"{u}ller, K. Z. Hatsagortsyan and C. H. Keitel, Phys. Rev. A \textbf{78}, 033408 (2008); C. Deneke and 
C. M\"{u}ller, Phys. Rev. A \textbf{78}, 033431 (2008).
\bibitem{Mueller_2009} C. M\"{u}ller, Phys. Lett. B \textbf{672}, 56 (2009).

\bibitem{Loetstedt_2008} E. L\"{o}tstedt et al. Phys. Rev. Lett. \textbf{101}, 203001 (2008); \textit{ibid.}, New J. Phys.
\textbf{11}, 013054 (2009).


\bibitem[{\citenamefont{Bulanov et~al.}(2006)\citenamefont{Bulanov, Narozhny,
  Mur, and Popov}}]{BuNaMuPo2006}
\bibinfo{author}{\bibfnamefont{S.~S.} \bibnamefont{Bulanov}} \bibnamefont{et~al.},
  \bibinfo{journal}{Zh. \'{E}ksp. Teor. Fiz.} \textbf{\bibinfo{volume}{129}},
  \bibinfo{pages}{14} (\bibinfo{year}{2006}), \bibinfo{note}{[JETP
  \textbf{102}, 9 (2006)]}.

\bibitem[{\citenamefont{Bell and Kirk}(2008)}]{bell:200403}
\bibinfo{author}{\bibfnamefont{A.~R.} \bibnamefont{Bell}} \bibnamefont{and}
  \bibinfo{author}{\bibfnamefont{J.~G.} \bibnamefont{Kirk}},
  \bibinfo{journal}{Phys. Rev. Lett.} \textbf{\bibinfo{volume}{101}},
  \bibinfo{eid}{200403} (\bibinfo{year}{2008}).

\bibitem[{\citenamefont{Ruf et~al.}(2009)\citenamefont{Ruf, Mocken, M\"{u}ller,
  Hatsagortsyan, and Keitel}}]{Rufetal2009}
\bibinfo{author}{\bibfnamefont{M.}~\bibnamefont{Ruf}} \bibnamefont{et~al.}, \bibinfo{journal}{Phys. Rev. Lett.}
  \textbf{\bibinfo{volume}{102}}, \bibinfo{eid}{080402}
   (\bibinfo{year}{2009}).

\bibitem{Hebenstreit_2009} F. Hebenstreit et al., Phys. Rev. Lett. \textbf{102}, 150404 (2009).

\bibitem[{\citenamefont{{Di Piazza}
  et~al.}(2008{\natexlab{a}})\citenamefont{{Di Piazza}, Hatsagortsyan, and
  Keitel}}]{DiPiHaKe2007}
\bibinfo{author}{\bibfnamefont{A.}~\bibnamefont{{Di Piazza}}},
  \bibinfo{author}{\bibfnamefont{K.~Z.} \bibnamefont{Hatsagortsyan}},
  \bibnamefont{and} \bibinfo{author}{\bibfnamefont{C.~H.}
  \bibnamefont{Keitel}}, \bibinfo{journal}{Phys. Rev. Lett.}
  \textbf{\bibinfo{volume}{100}}, \bibinfo{eid}{010403}
   (\bibinfo{year}{2008}{\natexlab{a}}).


\bibitem[{\citenamefont{{Di Piazza} et~al.}(2007)\citenamefont{{Di Piazza},
  Milstein, and Keitel}}]{DiPiMiKe2007}
\bibinfo{author}{\bibfnamefont{A.}~\bibnamefont{{Di Piazza}}},
  \bibinfo{author}{\bibfnamefont{A.~I.} \bibnamefont{Milstein}},
  \bibnamefont{and} \bibinfo{author}{\bibfnamefont{C.~H.}
  \bibnamefont{Keitel}}, \bibinfo{journal}{Phys. Rev. A}
  \textbf{\bibinfo{volume}{76}}, \bibinfo{eid}{032103}
   (\bibinfo{year}{2007}).

\bibitem[{\citenamefont{{Di Piazza} and Milstein}(2008)}]{DiPiMi2008}
\bibinfo{author}{\bibfnamefont{A.}~\bibnamefont{{Di Piazza}}} \bibnamefont{and}
  \bibinfo{author}{\bibfnamefont{A.~I.} \bibnamefont{Milstein}},
  \bibinfo{journal}{Phys. Rev. A} \textbf{\bibinfo{volume}{77}},
  \bibinfo{eid}{042102}  (\bibinfo{year}{2008}).

\bibitem[{\citenamefont{Mourou et~al.}(2006)\citenamefont{Mourou, Tajima, and
  Bulanov}}]{MoTaBu2006}
\bibinfo{author}{\bibfnamefont{G.~A.} \bibnamefont{Mourou}},
  \bibinfo{author}{\bibfnamefont{T.}~\bibnamefont{Tajima}}, \bibnamefont{and}
  \bibinfo{author}{\bibfnamefont{S.~V.} \bibnamefont{Bulanov}},
  \bibinfo{journal}{Rev. Mod. Phys.} \textbf{\bibinfo{volume}{78}},
  \bibinfo{eid}{309}  (\bibinfo{year}{2006}).

\bibitem{Marklund_2006} M. Marklund and P. K. Shukla, Rev. Mod. Phys. \textbf{78}, 591 (2006).

\bibitem[{\citenamefont{Salamin et~al.}(2006)\citenamefont{Salamin, Hu,
  Hatsagortsyan, and Keitel}}]{SaHuHaKe2006}
\bibinfo{author}{\bibfnamefont{Y.~I.} \bibnamefont{Salamin}} \bibnamefont{et~al.}, \bibinfo{journal}{Phys. Rep.}
  \textbf{\bibinfo{volume}{427}}, \bibinfo{pages}{41} (\bibinfo{year}{2006}).


\bibitem[{\citenamefont{Burke et~al.}(1997)\citenamefont{Burke, Field,
  Horton-Smith, Spencer, Walz, Berridge, Bugg, Shmakov, Weidemann, Bula
  et~al.}}]{Buetal1997}
\bibinfo{author}{\bibfnamefont{D.~L.} \bibnamefont{Burke}} \bibnamefont{et~al.},
  \bibinfo{journal}{Phys. Rev. Lett.} \textbf{\bibinfo{volume}{79}},
  \bibinfo{pages}{1626} (\bibinfo{year}{1997}).

\bibitem[{\citenamefont{Bamber et~al.}(1999)\citenamefont{Bamber, Boege,
  Koffas, Kotseroglou, Melissinos, Meyerhofer, Reis, Ragg, Bula, McDonald
  et~al.}}]{BaBoetal2004}
\bibinfo{author}{\bibfnamefont{C.}~\bibnamefont{Bamber}}
  \bibnamefont{et~al.}, \bibinfo{journal}{Phys. Rev. D}
  \textbf{\bibinfo{volume}{60}}, \bibinfo{pages}{092004}
  (\bibinfo{year}{1999}).

\bibitem[{\citenamefont{Cowan et~al.}(1999)\citenamefont{Cowan, Perry, Key,
  Ditmire, Hatchett, Henry, Moody, Moran, Pennington, Phillips
  et~al.}}]{CowPeretal1999}
\bibinfo{author}{\bibfnamefont{T.~E.} \bibnamefont{Cowan}}
  \bibnamefont{et~al.}, \bibinfo{journal}{Laser and Particle Beams}
  \textbf{\bibinfo{volume}{17}}, \bibinfo{pages}{773} (\bibinfo{year}{1999}).

\bibitem[{\citenamefont{Gahn et~al.}(2000)\citenamefont{Gahn, Tsakiris,
  Pretzler, Witte, Delfin, Wahlstr\"{o}m, and Habs}}]{gahnetal2000}
\bibinfo{author}{\bibfnamefont{C.}~\bibnamefont{Gahn}} \bibnamefont{et~al.},
  \bibinfo{journal}{Appl. Phys. Lett.} \textbf{\bibinfo{volume}{77}}
  \bibinfo{pages}{2662} (\bibinfo{year}{2000}).

\bibitem[{\citenamefont{Chen et~al.}(2009)\citenamefont{Chen, Wilks, Bonlie,
  Liang, Myatt, Price, Meyerhofer, and Beiersdorfer}}]{CheScoetal2009}
\bibinfo{author}{\bibfnamefont{H.}~\bibnamefont{Chen}} \bibnamefont{et~al.},
  \bibinfo{journal}{Phys. Rev. Lett.} \textbf{\bibinfo{volume}{102}},
  \bibinfo{eid}{105001} (\bibinfo{year}{2009}); C. M\"{u}ller and C. H. Keitel, Nature Photonics \textbf{3}, 245 (2009).

\bibitem{Yakovlev_1966} V. P. Yakovlev, Zh. Eksp. Teor. Fiz. \textbf{49}, 318 (1965) [Sov. Phys.
JETP \textbf{22}, 223 (1966)].

\bibitem{Di_Piazza_2009} A. Di Piazza et al., Phys. Rev. Lett. \textbf{103}, 170403 (2009).
\bibitem{Schuetzhold_2008} R. Sch\"{u}tzhold, H. Gies, and G. Dunne, Phys. Rev. Lett. \textbf{101}, 130404 (2008).

\bibitem{Emax} V. Yanovsky et al., Opt. Express {\bf 16}, 2109 (2008).


\bibitem[{\citenamefont{Ba{\u \i}er et~al.}(1975)\citenamefont{Ba{\u \i}er,
  Mil'ste{\u \i}n, and Strakhovenko}}]{BaMiSt1975}
\bibinfo{author}{\bibfnamefont{V.~N.} \bibnamefont{Ba{\u \i}er}},
  \bibinfo{author}{\bibfnamefont{A.~I.} \bibnamefont{Mil'ste{\u \i}n}},
  \bibnamefont{and} \bibinfo{author}{\bibfnamefont{V.~M.}
  \bibnamefont{Strakhovenko}}, \bibinfo{journal}{Zh. \'{E}ksp. Teor. Fiz.}
  \textbf{\bibinfo{volume}{69}}, \bibinfo{pages}{1893} (\bibinfo{year}{1975}),
  \bibinfo{note}{[Sov. Phys. JETP {\bf 42}, 961 (1976)]}.

\bibitem[{\citenamefont{Becker and Mitter}(1975)}]{BeMi1975}
\bibinfo{author}{\bibfnamefont{W.}~\bibnamefont{Becker}} \bibnamefont{and}
  \bibinfo{author}{\bibfnamefont{H.}~\bibnamefont{Mitter}},
  \bibinfo{journal}{J. Phys. A} \textbf{\bibinfo{volume}{8}},
  \bibinfo{pages}{1638} (\bibinfo{year}{1975}).

\bibitem{Landau_b_4} V. B. Berestetskii, E. M. Lifshitz, and L. P. Pitaevskii, \textit{Quantum Electrodynamics}
(Elsevier, Oxford, 1982), Sec. 99.


\bibitem{Baier_b_1994} V. N. Baier, V. M. Katkov, and V. M. Strakhovenko, \textit{Electromagnetic Processes at High Energies in Oriented Single Crystals}, (World Scientific Publishing Company, Singapore, 1994), Par. 7.4.
\bibitem{Baier_2005} V. N. Baier, V. M. Katkov and V. M. Strakhovenko, Sov. Phys. JETP \textbf{67}, 70 (1988); V. N. Baier and V. M. Katkov, Phys. Rep. \textbf{409}, 261 (2005).
\bibitem{Ritus_Review} V. I. Ritus, J. Sov. Laser Res. \textbf{6}, 497 (1985).

\bibitem{PFS} http://www.attoworld.de/research/PFS.html.
\bibitem{PDG} Particle Data Group: W.-M. Yao et al., J. Phys. G \textbf{33}, 1 (2006).

\bibitem{Gruener_2007} F. Gr\"{u}ner et al., Appl. Phys. B \textbf{86}, 431 (2007).

\end{thebibliography}
\end{document}